\documentclass[11pt,a4wide]{article}

\newcommand{\Vect}[1]{\ensuremath{\mathbf{#1}}}
\newcommand{\Tens}[1]{\ensuremath{\overline{\overline{#1}}}}

\newcommand{\Der}[2]{\ensuremath{\frac{\partial #1}{\partial #2}}}
\newcommand{\Dhyd}[2]{\ensuremath{\frac{D #1}{D #2}}}
\newcommand{\Grad}[1]{\ensuremath{\Vect{\nabla} #1}}
\newcommand{\Div}[1]{\ensuremath{\Vect{\nabla} \cdot #1}}

\newcommand{\R}{\ensuremath {\rho}}
\newcommand{\dR}{\ensuremath {\delta \rho}}
\newcommand{\dRk}{\ensuremath {{\delta \rho}_{\Vect{k}}}}
\newcommand{\dRkt}{\ensuremath { \delta \widetilde{\rho}}_{\Vect{k}}}

\newcommand{\V}{\ensuremath {\Vect{v}}}
\newcommand{\dV}{\ensuremath {\delta \V}}
\newcommand{\dVk}{\ensuremath {{\delta \V}_{\Vect{k}}}}
\newcommand{\dVkt}{\ensuremath { \delta \widetilde{\V}}_{\Vect{k}}}

\newcommand{\T}{\ensuremath {T}}
\newcommand{\dT}{\ensuremath {\delta T}}
\newcommand{\dTk}{\ensuremath {{\delta T}_{\Vect{k}}}}
\newcommand{\dTkt}{\ensuremath { \delta \widetilde{T}}_{\Vect{k}}}

\newcommand{\K}{\ensuremath {\Vect{k}}}
\newcommand{\X}{\ensuremath {\xi}}

\newcommand{\bibart}[6]{ #1, #2, \textbf{#3}, #4, (#5)}

\newcommand{\bibbook}[4]{ #1, \textit {#2}, #3, (#4)}

\begin{document}


\title{Quantitative study of a freely cooling granular medium}

\bigskip

\author{{\sc Pierre Deltour and Jean-Louis Barrat}\\
D\'epartement de Physique des Mat\'eriaux (UMR CNRS 5586)\\
Universit\'e Claude Bernard - Lyon I, 69622 Villeurbanne Cedex}

\date{\today}

\maketitle

\smallskip

Pacs numbers: {47.50.d, 05.20.Dd, 47.55.Kf}

\smallskip

Electronic mail: barrat@dpm.univ-lyon1.fr, pdeltour@dpm.univ-lyon1.fr

\bigskip

to be published in  Journal de Physique I : Statistical Physics

\bigskip

\begin{abstract}

We present a numerical study of a 
two dimensional granular medium consisting of hard inelastic disks. 
The evolution of the medium throughout a cooling process is
monitored. 
Two different types of instabilities (shearing and
clustering instability) are found to
develop in the system. The development of these instabilities
is shown to be in qualitative and quantitative agreement
with the predictions of linearized hydrodynamic theory.

\end{abstract}

\section{Introduction}
\label{intro}

Granular fluids are dense media  composed of elementary elements 
of  macroscopic size, undergoing collisions
in which their {\it macroscopic} energy is not conserved.
In the last decade, the flow of this granular fluids 
has received a great attention from the physics 
community, both because these media offer a "simple" 
example of dissipative systems and because of their 
numerous industrial applications. Two factors make the behaviour of 
granular fluids very different from that of molecular fluids. 
Firstly, the macroscopic size of the particles 
implies that external fields (boundaries or gravity) have a much
stronger effect on granular fluids.  Secondly, the energy of the 
granular fluid is not a conserved variable, since the heat 
dissipated in collisions can be considered as lost as 
far as the  flow is concerned. These two effects are often difficult
to disentangle in experiments, and also in  the numerical simulations
that aim at a realistic modelling of these experiments \cite{HERRMAN}.
A different type of numerical simulations was initiated   
by several groups \cite{Goldhirsch, Young1,Young3,Young2,Bernu}. In this approach
 external fields are  ignored, and only the dissipation is 
taken into account. This dissipation is moreover modelled in a
very simplistic way, by describing the particles as monodisperse
rigid hard spheres undergoing inelastic collisions. The dissipation is 
entirely specified by the restitution coefficient $r$, where $1-r$
is the 
fraction of the kinetic energy lost in a collision
(in the center of mass frame).

The aim of these simulations is not to model  actual experiments
involving granular fluids. Instead, the models are used to assess 
the difference in behaviour induced by the dissipation between a
granular fluid and its "atomic" counterpart ($r=1$, the hard sphere 
fluid.) In particular, the simulations can be used to investigate 
the validity of the hydrodynamic equations frequently used to 
describe granular flow  in more realistic situations 
\cite{Haff,Campbell}. As for atomic fluids, they also
provide a direct way of measuring the equation of state 
and transport coefficients that enter these equations. These 
transport
coefficients can then be compared to those obtained using
"granular kinetic theory" \cite{Jenkins}.

A particularly simple  and instructive situation that can easily
be studied in numerical simulations is the 2-dimensional
 "cooling problem"
first studied in  \cite{Goldhirsch, Young2}. In their simulations, 
these authors start from an equilibrium configuration of an
{\it elastic} ($r=1$) hard disc fluid. The behaviour of this fluid
after introducing a nonzero restitution coefficient is followed 
using  the standard molecular dynamics method for hard bodies
\cite{AT87}, with a collision rule that takes the 
dissipation into account. The kinetic temperature (average kinetic 
energy per particle) decreases due to the inelasticity of the
collisions, so that the fluid cools down as time increases.
It was shown in \cite{Goldhirsch, Young2} that, depending on the 
system size, on the restitution coefficient  and density, 
this cooling follows different routes. In the simplest case,
(small systems or small dissipation)
the fluid remains homogeneous at all temperatures. In larger 
systems or for larger dissipations, either the velocity 
field or the density field in the fluid develop 
instabilities and become inhomogeneous. It was also found by the 
same authors that the occurence of such instabilities is 
in qualitative agreement with the predictions of a linear stability 
analysis of the hydrodynamic equations for granular 
fluids \cite{Haff}. Finally, it was discovered in \cite{Young1,Young2}
that in some cases, the cooling ends at a finite time due to 
a singularity in the system dynamics, which was shown to correspond
to an infinite number of collisions within a finite time. This 
singularity, observed both in 1 and 2 dimensions, was described 
as an "inelastic collapse" of the system.  

In this paper, a detailed and quantitative analysis 
of the instability of homogeneous cooling of granular fluids 
is attempted. Our aim is to compare quantitatively the 
predictions of granular kinetic theory and granular hydrodynamics 
to the results of molecular dynamics simulations of the cooling 
of an inelastic hard disk fluid. The paper is organized as follows. 
The main predictions of granular hydrodynamics concerning the 
cooling problem are briefly recalled. Computational details 
concerning the simulation are given in section 3. The different 
regimes occuring during the cooling are analyzed in section
4, and compared with the theoretical predictions. Our main focus 
will be on the growth rate of the density instability, that 
can be computed by monitoring the structure factor
of the system as a function of time. Finally, the problem of the 
"inelastic collapse" 
is adressed in section 5, where a possible method for avoiding this 
singularity in the system dynamics is proposed.

\section{Hydrodynamic analysis of the cooling problem}

The hydrodynamic equations that have been 
proposed to describe granular flow \cite{Haff}
are based on mass and momentum conservation, and 
are very similar to the usual Navier-Stokes equations.  The only
modification is the appearance of a  new term in the energy 
(or temperature)
equation, accounting for the loss of energy in the collisions.
These equations can be compactly written in the 
form

\begin{eqnarray}
\Dhyd{\R}{t} & = & -\R \Div{ \V} \\
\R \Dhyd{\V}{t} & = & -\Div{\Tens{P}} \\
\R \Dhyd{T}{t} & = & -\Div{\Vect{Q}}-tr \left( \Tens{P}\,\Tens{D} \right )-\gamma T^\frac{3}{2}
\end{eqnarray}

where 
$D/Dt$ is the hydrodynamic derivative, $\Tens{D} $ the symmetrized
 velocity gradient tensor, \Tens{P} the stress tensor, \Vect{Q} the heat flux
and $\gamma$ represented the rate of energy lost due to inelastic collisions.
For a hard disk fluid, the equation of state
and the expression of the various transport coefficients
can be obtained from Jenkins and Richman kinetic theory \cite{Jenkins}.
These expressions are recalled in Appendix A.
The energy sink term,  $\gamma T^{3/2}$, has also been 
written in the form appropriate for hard disks. $\gamma$ in that case 
is a function of the density and the restitution coefficient,
which at least in the low density limit must
be proportionnal to $ \R $  and $(1-r)$. This can be understood
from the following reasoning: the kinetic energy loss
per particle per unit time is proportional to the collision
frequency (i.e. to $\R T^{1/2}$) and to the 
energy loss per collision $(1-r) T$.

A trivial solution of the cooling problem formulated above  corresponds
to an homogeneously cooling fluid, with a uniform density,
a vanishing velocity field, and a uniform temperature 
with an algebraic time decay
\begin{eqnarray}
\label{temperature}
T(t)=T_0{\left(1+\frac{t}{t_0}\right)}^{-2}
\end{eqnarray}.
Here $t_0=2 \R_0/(\gamma_0 {T_0}^{1/2})$ sets the time scale for temperature decay in the fluid.
The linear stability of this homogeneous solution
has been investigated in references \cite{Goldhirsch, McNamara}.
For completness, the main steps of this analysis
will be repeated here. The linearized equations
that describe the evolution of a sinusoidal
perturbation 

\begin{eqnarray}
\dR=\dRk \exp \left ( i \K \cdot r \right ) \\
\dV=\dVk \exp \left ( i \K \cdot r \right ) \\
\dT=\dTk \exp \left ( i \K \cdot r \right )  
\end{eqnarray}

around the homogeneous solution are
\begin{eqnarray}
\label{cont1}
\Der{\dRk}{t} &  = &  -i\R_0 \left ( \K \cdot \dVk \right ) \\
\label{eq:vitesseParral2}
\R_0 \Der{\K \cdot \dVk}{t} &  = &  -ik^2 \left [ \R_0 p'( \nu_0)\dTk+T_0 \left (p'(\nu_0)+\nu_0 \left ( \Der {p'}{\nu} \right )_0 \right )\dRk \right ]  \nonumber\\
 & &  \mbox{ } - \mu_0 \left [ k^2 \left ( \K \cdot \dVk \right ) \right ]  \\
\label{eq:vitessePerp2}
\R_0 \Der{\K_\perp \cdot \dVk}{t} &  = &  - \mu_0 \left [ k^2 \left ( \K_\perp \cdot \dVk \right ) \right ]   \\
\label{eq:temperature2}
\left [ \dRk \Der{T_0}{t}+\R_0 \Der{\dTk}{t} \right ] &  = &  -\kappa_0 k^2 \dTk
-{p_h}_0\left ( i \K \cdot \dVk \right) \nonumber\\
 & & \mbox{ } - \frac{3}{2} \gamma_0 {T_0}^\frac{1}{2} \dTk 
- T^{\frac{3}{2}}  \nu {\left( \Der{\gamma}{\nu} \right )}_0   \frac{\dRk}{\R_0} \end{eqnarray}
As for usual  fluids, the transverse part of the velocity
field completely decouples from the longitudinal part, and decays
with time as $ \left ( 1 +  {t}/{t_0} \right )^{-k^2 T_0^{1/2}
{t_0}/{\R_0}}
$. The longitudinal part of the velocity field,
the temperature and the density are coupled, and give rise
to three modes that have an algebraic time dependance
\begin {eqnarray}
\dRk & = & \dRkt { \left [ 1+ t/t_0 \right ]}^\X \\
\dVk & = & \dVkt  { \left [ 1+ t/t_0 \right ]}^{\X-1} \\
\dTk & = & \dTkt  { \left [ 1+ t/t_0 \right ]}^{\X-2} 
\end{eqnarray}.
The exponents $\xi(k)$ for the three modes are the three
roots of the determinant  of the following set of equations
\begin{eqnarray*}
\dRkt \left (\frac{\X(\X-1)}{t_0^2}+ \K^2 T_0 \left [p'(\nu_0)+\nu_0 \left ( \Der {p'}{\nu} \right )_0 \right ] +  \frac{\mu_0 k^2} {\R_0 t_0} \X \right )
+ \dTkt \left [ k^2 \R_0 p'(\nu_0) \right ] & = &  0 \\
\dRkt \left [  \frac{\T_0}{t_0} \left (-2 
+ T_0^{\frac{1}{2}} \nu {\left( \Der{\gamma}{\nu}\right )}_0 \frac{t_0}{\R_0}
- p'(\nu_0) \X \right ) \right ] 
+ \dTkt \left [ (\X+1) \frac{\R_0}{t_0} + \kappa_0 k^2 \right ] &  = & 0
\end{eqnarray*}
A typical plot of the wavevector dependance
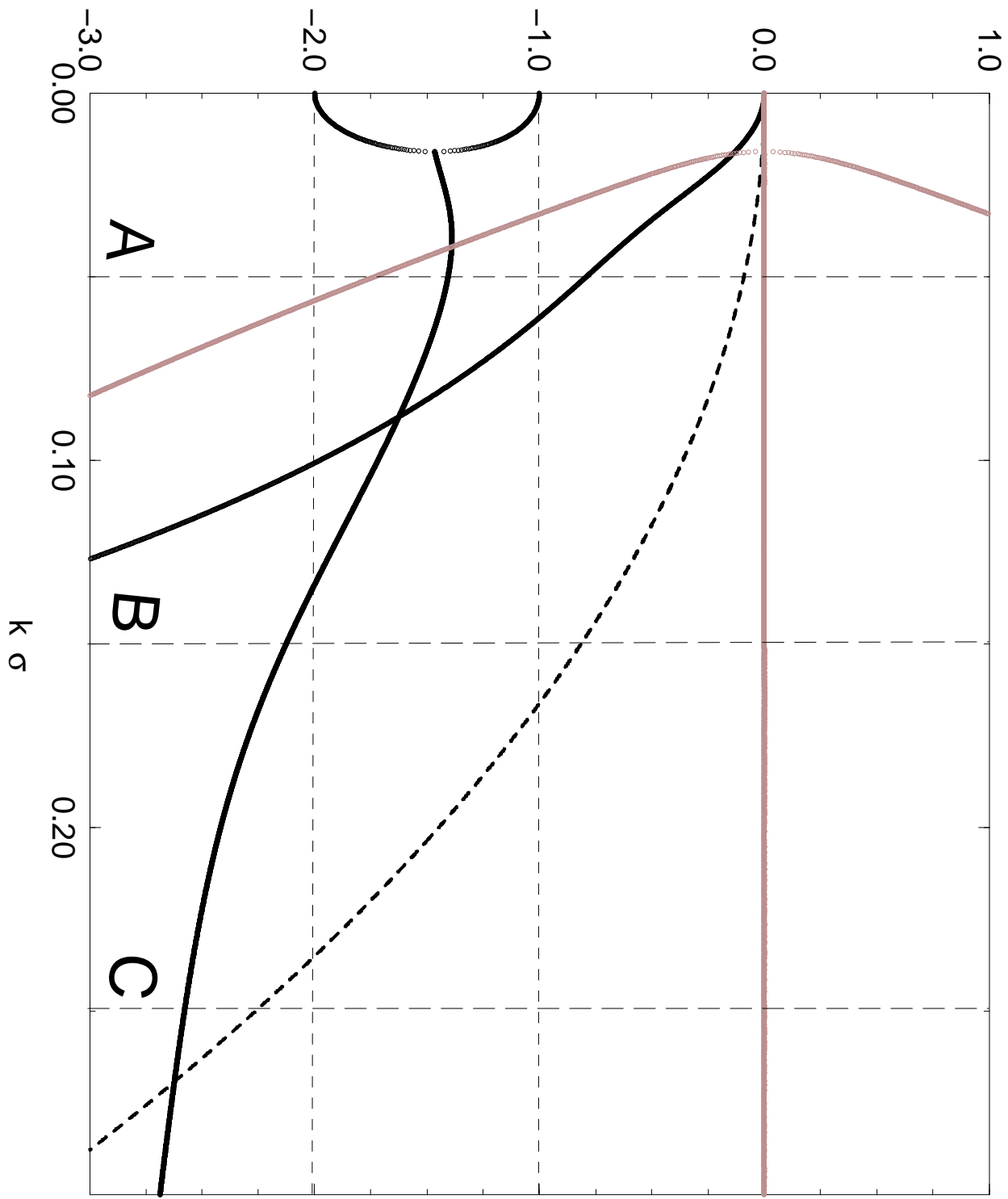
of these three roots, together with the growth rate 
of the  velocity perturbations, is shown in figure 1.
It must be emphasized that the stability of velocity
disturbances is determined by the comparison
between the growth exponent  of the disturbance with
the value $-1$ that characterizes the decay of the 
{\it thermal} velocity. Hence a growth exponent larger than
$-1$ for the transverse or longitudinal velocity
fields is indicative of an instability of the macroscopic
velocity. If the growth exponent of the longitudinal 
velocity field is larger than $-1$, a corresponding instability 
in the density field will follow from equation \ref{cont1}.

This analysis yields to the prediction of three different 
possible behaviours of the system, depending on the 
value of the parameters and on the system size, that 
introduces a lower wavevector cutoff. If this lower cutoff 
corresponds to the line $C$ of figure 1, the homogeneous solution
will be linearly stable. This regime will be described
as the homogeneous kinetic regime. If the lower cutoff moves 
to the abscissa indicated by line $B$ in figure 1, the transverse 
velocity field will become unstable while the system remains 
homogeneous. In this "shearing" regime, first observed in
reference \cite{Goldhirsch}, a shearing flow will 
develop in the system.  Finally, for a lower cutoff
corresponding to abscissa $A$, an instability of 
the longitudinal velocity field and the corresponding
instability in the density field will take
place together with the shearing instability. 
 In this "clustering" regime, the growth
of density disturbances will yield to the formation
of dense clusters of particles, as first observed in
reference \cite{Hopkins}.

All three situations have already been observed
in numerical simulations of the cooling in two dimensional
granular fluids. The aim of the next sections
will be to attempt a quantitative analysis of the 
behaviour of a cooling granular fluid, and to compare the results
to the predictions summarized above.

\section{Computational details}

The model simulated in this work is in all respects similar
to that studied in \cite{Goldhirsch, Young2}. The system is made up of
$N$ hard inelastic disks of diameter $\sigma$, in a square cell of
size $L$ with periodic boundary conditions. The cell size
$L$ sets the lower cutoff in wavevector space, $k_{min}=2\pi/L$.
A standard cell-linked Molecular Dynamics algorithm for hard bodies 
\cite{AT87} is used. In a first step,  the system is equilibrated 
with a  coefficient $r$ equal to unity. At time
$t=0$, inelasticity is switched on  and cooling starts,
with an initial
temperature $T_0$. The restitution coefficient
enters through a simple modification of 
the standard collision rule between hard disks, the velocities
of the two disks after a collision being given by
\begin{eqnarray}
\label{coll1}
\Vect {u_1}'=\Vect{u_1}-\frac{1}{2}(1+r)
[\hat{\Vect{n}} \cdot (\Vect{ u_1}-\Vect {u_2}) ] \hat{\Vect{n}} \\
\label{coll2}
\Vect {u_2}'=\Vect{u_2}+\frac{1}{2}(1+r)
[\hat{\Vect{n}} \cdot (\Vect{ u_1}-\Vect {u_2}) ] \hat{\Vect{n}} 
\end{eqnarray}
where the primes denote the quantities after collision and $ \hat{{\Vect{n}}} $
is a unit vector along the centers line from particle 1 towards particle 2.
The natural units in this problem are the particle mass $m$
 and diameter,
and the thermal energy at $t=0$,
 i.e. $T_0$. The corresponding time unit
is $\tau=(m/T)^{1/2}\sigma$. The state of the system
is defined by three dimensionless numbers,
which are the reduced size $L/\sigma$ (or equivalently
the reduced cutoff $k_{min}^*= k_{min}\sigma$), the reduced density
$\rho*=\sigma^2 N/L^2 $, and the restitution coefficient $r$.

The state of the fluid during the cooling was
monitored by a systematic computation
of coarse-grained (hydrodynamic) density and velocity fields.
The coarse graining is obtaining here from a division of
the system into 100 square subcells. Besides, statistical quantities
characterizing the state of the system have 
also been systematically computed. These quantities are
the momenta of the velocity distribution of 
individual particles, the pair correlation function $g(r)$
for interparticle distance, and the structure factor
\begin{eqnarray}
S(\Vect{k}) & = &  \frac{1}{N}  \rho_\Vect{k} \rho_\Vect{-k} 
\end{eqnarray}
This structure factor can be computed
for all wavevectors compatible with 
the periodic boundary condition, of the form $(n_x,n_y) k_{min}$.
As the system is not in a stationary state, these 
quantities are time dependant. A large enough system 
is thus necessary to obtain reasonable statistics
without time averaging. The values of $N$ investigated in this work
vary from $N=1600$ to $N=10000$.

\section{results}

 \subsection{Kinetic regime}

According to the analysis of section \ref{intro},
the kinetic regime corresponding to 
a stable homogeneous cooling will be observed 
(at a given density and restitution coefficient) 
for small enough systems. Such a situation allows 
a clear testing of  some of the hypothesis 
of the kinetic theory  description of the granular fluid.
In particular, the pair correlation function and
velocity distribution can be compared to that of 
an elastic hard disk fluid throughout the cooling 
process. The temperature decay can be monitored 
and compared to the theoretical prediction
(equation \ref{temperature}), and the decay time $t_0$ (or equivalently the 
coefficient $\gamma(\rho)$) compared to the prediction
of kinetic theory. 

The pair correlation of an  homogeneously 
cooling granular fluid after the temperature
has dropped by a factor of 10 is shown in figure 2. This comparison
shows that the local structure of the cooling
granular medium (which determines
its equation of state) remains essentially identical to that 
of an equilibrium fluid.  The study of the 
velocity distribution function shows that this distribution remains
maxwellian throughout the cooling. 

This similarity between the 
structure and velocity distribution of the 
granular fluid and the usual hard disk fluid 
suggests that the kinetic theory of Jenkins \cite{Jenkins} is applicable.
This expectation is borne out by the study of the time dependance of
the fluid temperature. As shown in figure 3, the temperature decay
is perfectly described by equation \ref{temperature}. The density dependance
of the decay time $t_0$ is compared in figure 4  to the prediction
of kinetic theory (see appendix B). The agreement is extremely good,
and suggests that all the transport 
coefficients appearing in the hydrodynamic
equations can be estimated using this kinetic theory. 

 \subsection{Shearing regime}

If the restitution coefficient $r$ decreases 
or if the size of the system increases, the hydrodynamic theory
predicts a regime in which transverse fluctuations of
the velocity field are unstable. This regime is indeed
observed in the simulations, as shown in figure 5.
A shear flow that corresponds to the smallest wavevector 
compatible with the periodic boundary conditions develops
in the system. In this regime, the total kinetic energy of
the system (which in that case is not the temperature, since
the system has developped an ordered flow pattern) appreciably deviates
from equation \ref{temperature}, as shown in figure 6.

 \subsection{Clustered regime}

For even larger systems 
or smaller restitution coefficients,  the cooling granular fluid
becomes inhomogeneous, as shown in figure 7.
this  spontaneous  formation of density inhomogeneities
(or clusters) was first observed in the simulations of the cooling problem
by Goldhirsch and Zanetti and Young and McNamara \cite{Goldhirsch, Young2}.
Two different explanations have been put forward to explain this
cluster formation. The first one, found in \cite{Goldhirsch}, is to consider this
cluster formation as a secondary instability of the shearing regime,
due to the developpment of temperature and pressure gradients 
in the shearing regime.  The second possible explanation is 
that cluster formation is directly related to the linear instability of the 
density modes predicted by hydrodynamic theory. 

 In order to characterize 
quantitatively this clustering regime, the structure factor 
$S(k,t)$ of the system has been computed as a function of time 
and wavevector. The corresponding data is shown in figure 
8. The growth of the  density inhomogeneities results
in the appearance of a low wavevector peak in the
structure factor, that rapidly increases with time. 
According to hydrodynamics, the time dependance 
of $S(k,t)$ should be algebraic, i.e.
\begin{eqnarray}
S(\K,t) & = & S(\K,0) {\left ( 1 + \frac{t}{t_0} \right )}^{2\xi(\K)}
\end{eqnarray}
so that the ratio
\begin{eqnarray}
\frac{\ln (S(\K,t))-\ln (S(\K,0))}{ln \left ( 1 + \frac{t}{t_0} \right )} & =
 & 2\xi(\K)
\end{eqnarray}
should be independent of time. This ratio is plotted in figure 9  as 
a function of wavevector for different times.  $2\xi(k)$ seems to be
reasonably independent of time, and  its low wavevector 
value appears to be consistent with the prediction of linearized 
hydrodynamics.  Hence the  density instability
can be interpreted as resulting from 
a linear instability of the homogeneous solution
of the hydrodynamic equations. Note that it was recently observed
by McNamara and Young that the "clustering" fluid
eventually develops for long times into an ordered flow pattern
of the "shearing" type. This is also consistent with hydrodynamics,
since the growth rate of the transverse velocity modes is 
positive. The description of the formation of this shearing flow
in an inhomogeneous system, however, is beyond the possibilities
of linearized hydrodynamics.

\section{Inelastic collapse and how to avoid it}

The inelastic collapse singularity was first observed
by \cite{Bernu, Young1} in simulations of unidimensional
inelastic system. This collapse can be described as the 
appearance of an infinite number of correlated collisions between
a few particles, taking place in a finite time. The same phenomenon 
was observed in two dimensions by \cite{Young2}. It was shown 
that in that case the correlated collisions take place
between a small number of essentially aligned particles, 
so that the unidimensional situation is practically 
reproduced.

In order to avoid this inelastic collapse, a slightly 
modified collision rule between the particles can be introduced.
At each collision, the relative velocity 
of the two particles is first computed according to the 
usual rule (equations \ref{coll1} and \ref{coll2}), then rotated by a small 
(less than 5 degrees) random angle.
This can be justified by invoking the unavoidable
roughness of actual solid particles, conservation 
of angular momentum being (virtually) ensured by a transfer to
the internal degrees of freedom of the particles.
As to inelastic collapse, the aim of this modified collision
rule is to hinder the formation of correlated particle lines 
that cause this singularity.  Indeed, inelastic collapse
was not observed in the simulations where this "random"
collision rule was used, while under the same conditions 
a system following the "deterministic" collision 
rule always underwent inelastic collapse (figure 10).
Hence inelastic collapse appears to be a pathology related to the
use of purely specular collision rule
between particles, rather than a characteristic of inelastic 
fluids.

\section{Conclusion and perspectives}

The main objective of this work
was to assess the validity of the hydrodynamic 
description of granular fluids originally proposed
by \cite{Haff}, and of the  kinetic theory calculation of
the associated transport coefficients. The study of the 
particularly simple  "cooling fluid" case 
and of the associated instabilities provides an ideal
benchmark for this description. The comparison between numerical
simulations and theoretical predictions in this simple case
shows that the  theory is quantitatively accurate.
A similar  conclusion was also reached in a recent study by McNamara and
Young \cite{preprint}, who showed that the transitions between the
different cooling regimes were correctly predicted by the theory.

The description of the inelastic collapse phenomenon observed
by McNamara and Young is obviously beyond
the possibilities of kinetic theory or hydrodynamics. It was
shown that this phenmenon can easily be avoided by introducing a small amount of
randomness in the collisions between particles, similar to what would
be caused by the natural roughness of granular particles.

Obviously, a correct description of granular fluid 
cannot be achieved without a knowledge of the boundary conditions 
that must be used for  the hydrodynamic equations. These conditions, and
in particular those that correspond to the very important case of vibrating 
solid walls, are not known. Their determination, through the quantitative
comparison of numerical simulation and theory, will be the subject of future 
work.

\section*{Ackowledgments}

This work was supported by the Pole Scientifique de Mod\'elisation
Num\'erique at ENS-Lyon.


\pagebreak
\appendix

\section{Expressions for the transport coefficients and the equation of state}

The Navier-Stokes like equations describing a granular fluid are:

\begin{eqnarray*}
\label{eq:masse1}
\Dhyd{\R}{t} & = & -\R \Div{ \V} \\
\label{eq:vitesse1}
\R \Dhyd{\V}{t} & = & -\Div{\Tens{P}} \\
\label{eq:temperature1}
\R \Dhyd{T}{t} & = & -\Div{\Vect{Q}}-tr \left( \Tens{P}\,\Tens{D} \right )-\gamma T^\frac{3}{2}
\end{eqnarray*}

$D/Dt$ is the hydrodynamic derivative, $\Tens{D} $ the symmetrized
 velocity gradient tensor, \Tens{P} the stress tensor, \Vect{Q} the heat flux
and $\gamma$ represents the rate of energy lost due to inelastic collisions. 
The definition for these quantities is:

\begin{eqnarray*}
\Dhyd{ }{t} & = & \Der{}{t}+\left( \V \cdot \Grad{} \right ) \\
D_{ij} & = & \frac{1}{2} \left( \Der{v_i}{x_j}+\Der{v_j}{x_i} \right )\\
\Tens{P} & = & p_h \Tens{1}-2\mu \left ( \Tens{D}-\frac{1}{2} \left( \Div{\V} 
\right ) \Tens{1} \right)\\
\Vect{Q} & = & -\kappa \Grad{T}
\end{eqnarray*}

The various  transport coefficients and equation of state are 

\begin{eqnarray*}
p_h & = & p'(\nu) \R T  \\
\mu & = & \mu'(\nu) T^\frac{1}{2} \R \sigma \\
\kappa & = & \kappa'(\nu) T^\frac{1}{2} \R \sigma \\
\gamma & = & \gamma' (\nu) \frac{\R}{\sigma}
\end{eqnarray*}

  where $p'$, $\mu'$, $\kappa'$, $\gamma'$ are functions only
of the solid fraction $\nu=\R/\R_s$.

\begin{eqnarray*}
p'(\nu) & = & \frac{2\nu+s_{*}}{s_{*}} \\
\mu'(\nu) & = & \left ( \frac{\nu^2}{\sqrt{\pi}}+ \frac{\sqrt{\pi}}{8} \left ( \nu + s_{*} \right ) ^{2} \right ) \frac {1}{\nu s_{*}} \\
\kappa'(\nu) & = & \left ( \frac{1}{\sqrt{\pi}}+\frac{\sqrt{\pi}}{2} \left ( \frac{3}{2}\nu + s_{*} \right )^{2} \right ) \frac{1}{\nu s_{*}} \\
\gamma'(\nu) & = & \frac{8}{\sqrt{\pi}} \left ( 1-r \right ) \frac{\nu}{s_{*}} 
\end{eqnarray*}

where $s_{*}$ is defined as

\begin{eqnarray*}
s_{*}(\nu) & = & \frac{\left(1-\nu \right )^{2}}{\left ( 1 - 7\nu/16  \right )}
\end{eqnarray*}

\section{Enskog expansion}

In the case of a hard core fluid, a semi-empirical modification of Boltzmann
equation introduced by Enskog, widens the range of applicabity of the
kinetic approach to higher densities \cite{Resibois et De Leener}.
 The Enskog approximation accounts for 
the finite size of the disks  in the collision term
of the Boltzmann equation. When  two particles collide, their centers
are separated by the diameter of the disks  $\sigma $. The collision
term of Boltzmann equation should thus be multiplied by the probability
of finding two particles separated by $\sigma$ which is  proportional
to the pair correlation function evaluated at $\sigma$. This correction
will have an influence on all the transport coefficients. The validity  of this approach  was checked for the cooling rate $\gamma$ in the kinetic regime.

Figure 2 shows successive snapshots of the pair correlation function in the
kinetic regime. This function is essentially the same as in  a hard core fluid at thermodynamic equilibrium even though the temperature has dropped by a factor 10 between the first and the last snapshot.  Hence  $g(\sigma)$ is assumed to be given by the usual
virial expression
\begin{eqnarray*}
\frac{p_h}{\R T} & = & 1+ 2 \nu g(\sigma)
\end{eqnarray*}

Introducing the equation of state of an 2d hard disks fluid
\cite{Henderson:eqOfState}, the  Enskog 
corrected cooling rate $t_0$ becomes:
\begin{eqnarray*}
 t_0  & = & {t_0}_{\mbox{\tiny Boltzmann}} \frac{1}{g(\sigma)} \\
 & = & \left( \frac{1}{\frac{4}{\sqrt{\pi}}(1-r)\nu} \frac{\sigma}{{T_{0}}^{1/2}} \right )
\frac{\left( 1- \nu \right )^2}{(1-7\nu/16)} 
\end{eqnarray*}

The values of the cooling rate found in the simulation are compared 
in figure 4 with  this  prediction.

\pagebreak



{\bf figure 1:}
The decay rate of the velocity field 
disturbances computed by solving equation (4) for $\rho=0.2, r=0.9$
is shown in this figure as a function of wavevector. The dashed line indicates the decay rate of the
transverse modes. The black and grey lines correspond  respectively to  the real and imaginary  parts of the decay rate of the 
three longitudinal modes.

 {\bf figure 2:} Pair correlation functions g(r) computed for the initial (in black shifted up by 0.5)
and final (in grey) configurations in an homogeneously cooling system 
($N=1600, d=0.8, r=0.98$). The temperature drops
by a factor of ten without changes in the pair correlation function.

{\bf figure 3:} Evolution of the square root of the inverse  temperature 
versus t
in an homogeneously cooling system ($r=0.99, d=0.1, N=1600$). The 
solid line corresponds to the hydrodynamic prediction in the kinetic regime.

{\bf figure 4:}
 Enskog corrected value of the decay time calculated as a function
of density for $r=0.98$, compared to the values obtained in the simulations ($N=1600$).

{\bf figure 5:} Velocity field in the shearing regime after the granular medium has 
spontaneously developped a flow pattern corresponding to  the lowest wave 
vector compatible with the boundary conditions. (N=1600, d=0.1, r=0.92)

I{\bf figure 6}: square root of the inverse of the temperature versus t 
in the shearing regime. 
The solid line extrapolates towards  the  first moments of the run. There is a 
substantial deviation from this kinetic regime fit. ($N=1600, d=0.1, r=0.92$)

{\bf figure 7 } Final configuration (141 collisions per particle) of a simulation 
 in the cluster regime. ($N=1600, d=0.25, r=0.6$).

{\bf figure 8:} Evolution of the structure factor during the clusters formation 
($N=1600, d=0.5, r=0.4$).  The curves from the bottom
are separated by 10 collisions per particle. 
 Note the large increase of the structure factor in the long wavelength limit ($k \rightarrow 0$).

{\bf figure 9:}  Growth exponent of the density field disturbance, obtained from the simulation using equation (19)
 (N=10000, d=0.5, r=0.9)
The different symbols correspond to different times. The
 prediction of the hydrodynamic description of the instability is 
the solid line. 

{\bf figure 10} A system with $N=1600,d=0.25, r=0.25$ obeying  the specular collision rule collapses 
after 3.77 collisions per particle. The grey particles are those 
involved in the last two hundred collisions.
 The aligned particles represent more than 99 \% 
of the grey particles. Under the same conditions, a system obeying
the modified collision rule does not undergo collapse after 125 collisions per
particle.

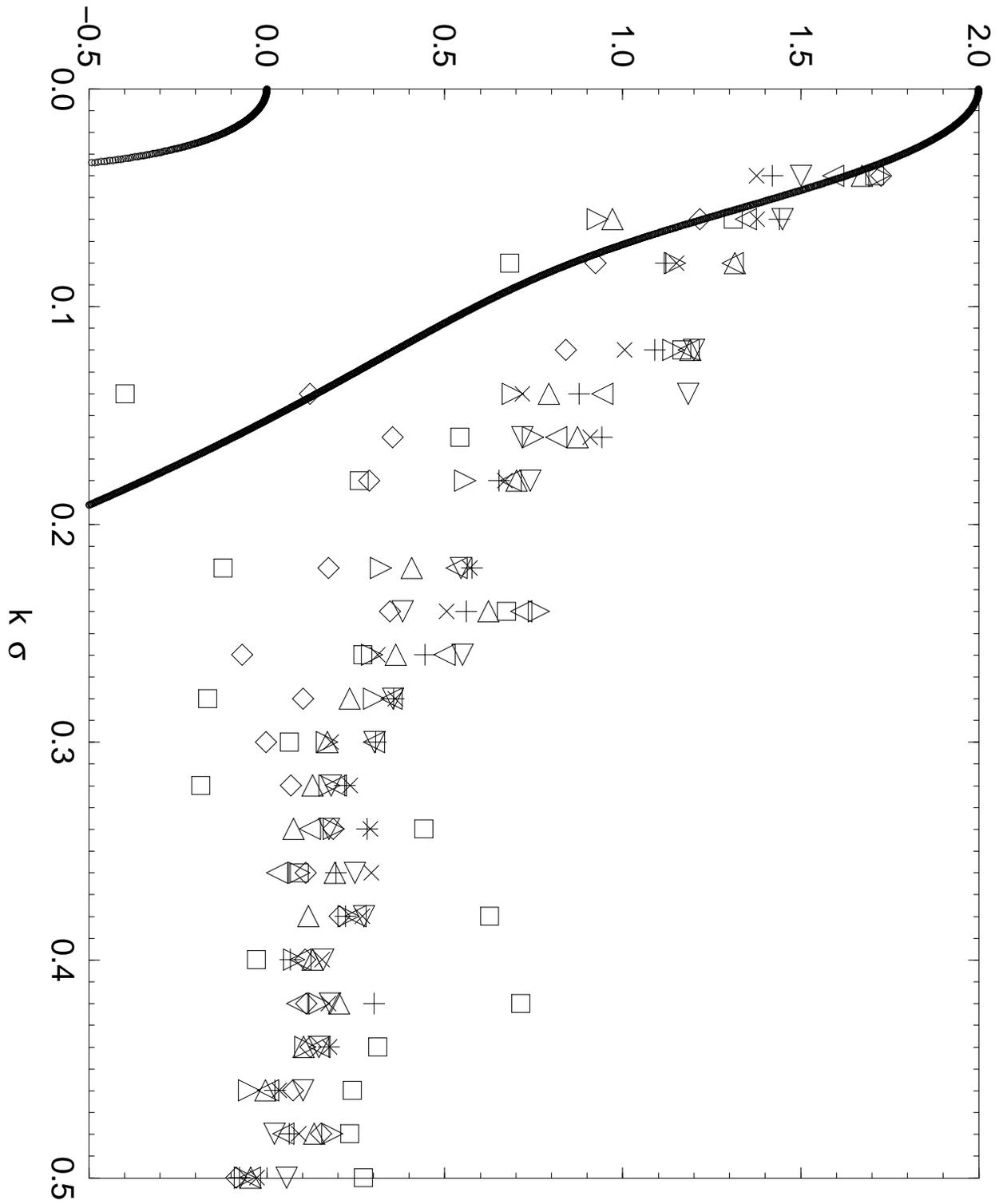
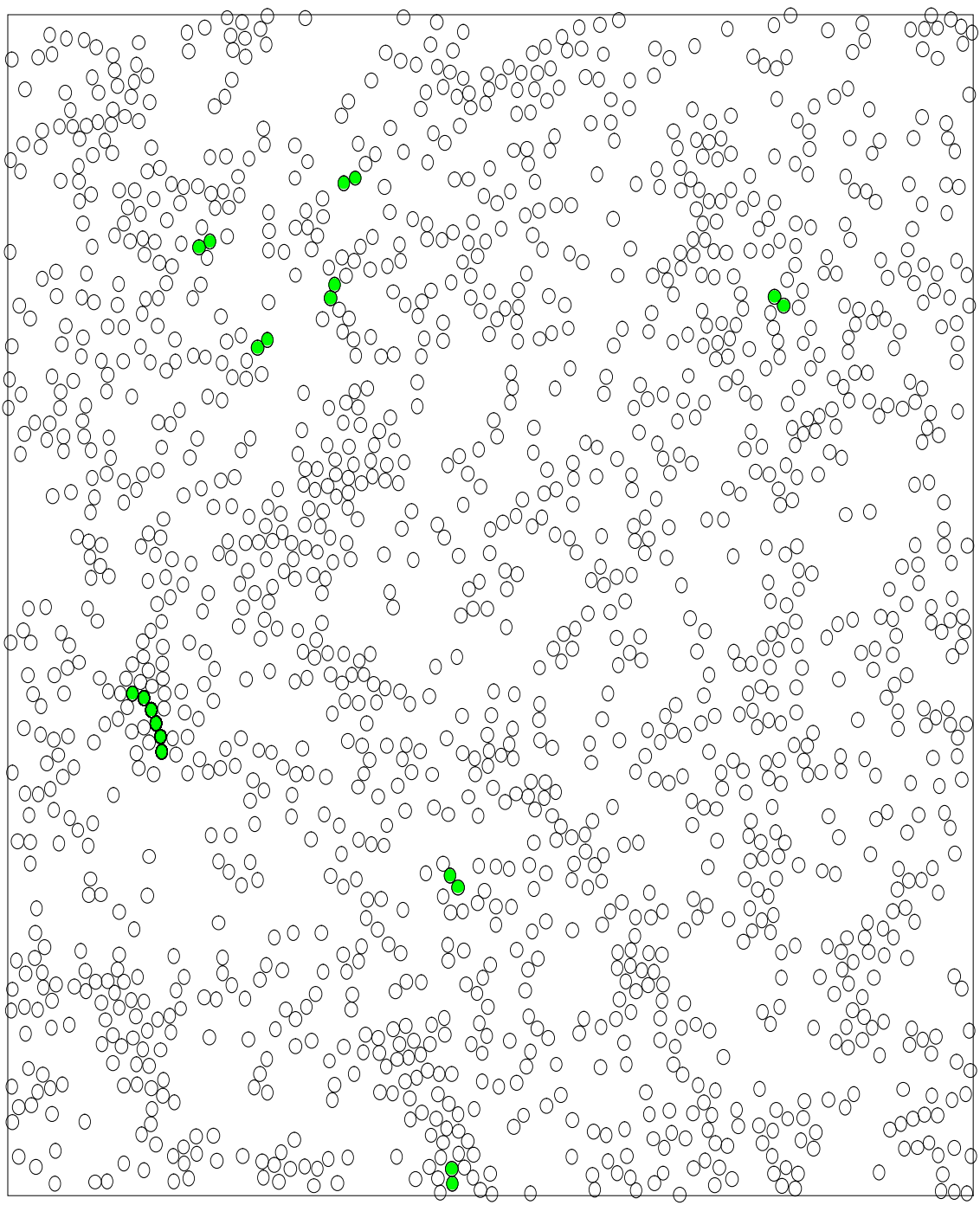
\end{document}